\documentclass[10pt,aps,twocolumn,longbibliography, notitlepage,showpacs,superscriptaddress, 
]{revtex4-2}
\usepackage{graphicx}
\usepackage{amssymb,amsmath,bm}
\usepackage{footnote}
\usepackage[colorlinks, linkcolor=blue,anchorcolor=blue,citecolor=blue,urlcolor=blue]{hyperref}
\usepackage{comment}
\usepackage{lipsum}  
\usepackage{siunitx}
\usepackage{xcolor}

\newcommand{\be}{\begin{equation}}
\newcommand{\ee}{\end{equation}}

\renewcommand{\b}[1]{{\boldsymbol{#1}}}

\begin{document}
\title{Topological Interlayer Superconductivity in a van der Waals Heterostructure}

\author{Even Thingstad}
\affiliation{Department of Physics, University of Basel, Klingelbergstrasse 82, CH-4056 Basel, Switzerland} 

\author{Joel Hutchinson}
\affiliation{Department of Physics, University of Basel, Klingelbergstrasse 82, CH-4056 Basel, Switzerland} 

\author{Daniel Loss}
\affiliation{Department of Physics, University of Basel, Klingelbergstrasse 82, CH-4056 Basel, Switzerland} 

\author{Jelena Klinovaja}
\affiliation{Department of Physics, University of Basel, Klingelbergstrasse 82, CH-4056 Basel, Switzerland} 
\date{\today}
\begin{abstract}
We show that when a honeycomb antiferromagnetic insulator (AFMI) is sandwiched between two transition metal dichalcogenide (TMD) monolayers in a commensurate way, magnons in the AFMI can mediate an interaction between electrons in the TMDs that gives rise to interlayer Cooper pairing. This interaction opens coexisting extended $s$-wave and chiral $p$-wave superconducting gaps in the energy spectrum of the coupled system, and the latter give rise to topological Majorana edge modes. 

\end{abstract}
\maketitle


\emph{Introduction.---}
The last decade has seen a proliferation in the number and variety of few-layer van der Waals heterostructures, which provide new grounds to study exotic many-body phases and their applications in quantum technologies~\cite{geim2013, conti2020, terrones2013}. 
Transition metal dichalcogenides (TMDs) are a particularly interesting class of such materials due to the interplay of their spin and valley degrees of freedom~\cite{Manzeli2017_2DTransitionMetal}. This interplay can give rise to strongly correlated phases with charge density wave, magnetic, and excitonic orders~\cite{chen2016, Chen2018_Reproduction, miserev2019, Arora2019_ExcitedState, Wang2020_CorrelatedElectronicPhases, Regan2020_MottGeneralizedWigner, Ma2021_StronglyCorrelatedExcitonic, Pan2022_Topological, Xu2022_InteractionDriven, Kiese2022_TMDs, Dong2023_CompositeFL, Roch2020_FirstOrder}. 
Furthermore, superconductivity has been observed in several TMD monolayers at large electron doping~\cite{Ye2012_SuperconductingDomeGateTuned, Lu2015_Evidence, Saito2016_Superconductivity, Xi2016, delaBarrera2018, Lu2018_Full, ding2022}, 
and topological superconductivity has been predicted~\cite{Yuan2014_Possible, hsu2017, he2018, Wang2018_Platform, Shaffer2020_Crystalline, Scherer2022_ChiralSuperconductivityEnhanced, Crepel2023_Topological, Zerba2023_Realizing}.

In multilayer systems, the additional layer degree of freedom enables further instabilities~\cite{hutchinson2024, Zhao2023_EvidenceFinitemomentumPairing}. 
An exotic possibility is interlayer superconductivity, where Cooper pairs are formed from electrons in different layers~\cite{Hosseini2012_ModelExoticChiral, Hosseini2012_Unconventional, Liu2017_Unconventional, Alidoust2019_Symmetry, Gu2023_EffectiveModelPairing}. This phenomenon could be exploited for an efficient Cooper-pair-splitting device 
~\cite{recher2001,Hofstetter2009_CooperPairSplitter, Herrmann2010_Carbon, Schindele2012_NearUnity, deacon2015}.
Current devices typically filter the individual electrons of Cooper pairs through spin-polarized quantum dots~\cite{bordoloi2022, Wang2022_SingletTripletCooper}. However, this process is limited by the inability to fully spin-polarize the dots.
With interlayer pairing, this issue could be circumvented by filtration in real-space.

\begin{figure}[tbp]
\includegraphics[width=0.98\columnwidth]{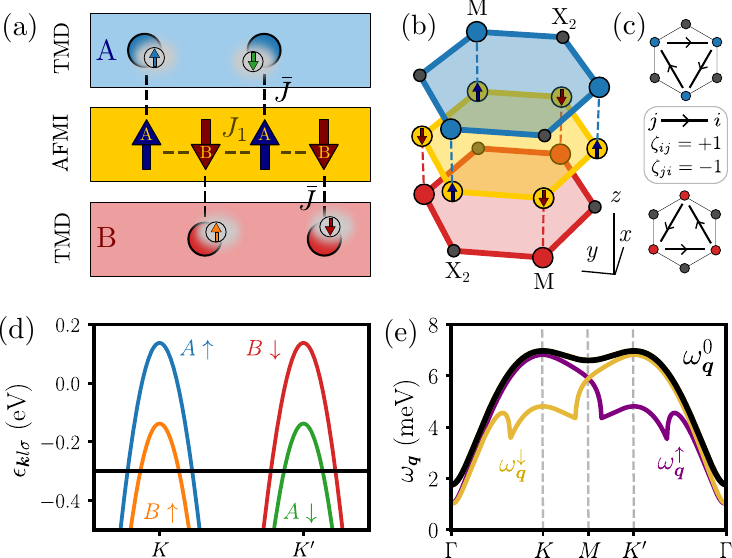}
\caption{ (a-b) Model heterostructure consisting of an AFMI sandwiched between two TMDs (labelled as $A$ and $B$). The electrons of TMD A (B) are coupled to the localized spins on sublattice A (B) of the AFMI.  The TMDs are characterized by strong Ising spin-orbit coupling that gives rise to a spin- and direction-dependent hopping \(t_{ij}^\sigma = t_0 + i \sigma \zeta_{ij} t\) with sign determined by \(\zeta_{ij}\) as shown in the panel (c). (d) The TMD energy spectrum with interlayer exchange interaction \(\bar{J} = \SI{55}{\milli\electronvolt}\).
 (e) The bare (black) and renormalized (yellow and purple) magnon spectra for nearest neighbour interactions. 
}

\label{fig:heterostruct}
\end{figure}

One mechanism for superconductivity in heterostructures is spin fluctuations in adjacent magnetic insulators~\cite{Rohling2018_Superconductivity, Fjaerbu2019_Superconductivity, Erlandsen2019_Enhancement, Thingstad2021_Eliashberg, Brekke2023_Twisted, Kargarian2016_Amperean, Hugdal2018_Magnon, Erlandsen2020_MagnonMediated, Bostrom2023_Topological, Sun2024_Strong}. Furthermore, the magnons of non-collinear magnetic textures can mediate topological superconductivity~\cite{Maeland2023_Skyrmionic, Maeland2023_Helical}.
Heterostructures containing magnets coupled to TMDs already provide fertile ground for new physics~\cite{sierra2021, bora2021, pei2019, Glodzik2020_Engineering}. In this paper, we uncover a mechanism for interlayer superconductivity in such a heterostructure: two TMD monolayers separated by an antiferromagnetic insulator (AFMI), see Fig.~\ref{fig:heterostruct}.
We show that magnons in the AFMI induce superconductivity with coexisting chiral $p$-wave and $s$-wave pairing. The topological nature of this state results in Majorana modes delocalized over the edges of the TMD bilayer.

\emph{Facilitating interlayer pairing.---}
Interlayer Cooper pairing requires Fermi surfaces engineered with a favourable phase space for the appropriate pairing channel, as well as interactions inducing this pairing. We focus exclusively on the case of zero-total-momentum pairs where the spin index $\sigma$ (quantization axis perpendicular to the layer planes) and layer index $l$ 
remain good quantum numbers~\footnote{The latter requires an insulating barrier to suppress layer hybridization.}.
To disfavor competing intralayer Bardeen-Cooper-Schrieffer (BCS) pairs, the normal state should have broken time-reversal symmetry \(\mathcal{T}\). In terms of the single-particle spectra \(\epsilon_{\bm{k}l\sigma}\), this implies $\epsilon_{\b{k}l\sigma} \neq \epsilon_{-\b{k} l \bar{\sigma}}$.  To favour interlayer pairing, the system should also preserve one out of two possible discrete interlayer symmetries, $\epsilon_{\b{k} l\sigma}= \epsilon_{-\b{k}, \bar{l},\pm\sigma}$, where the plus sign in the subscript corresponds to preserved inversion symmetry $\mathcal{I}$, and the minus sign to preserved symmetry $\mathcal{T}\otimes\sigma_h$, where $\sigma_h$ represents mirror reflection in the layer plane.
Since low-dimensional systems have reduced Coulomb screening, we expect the density-density interaction channel to be repulsive and look for unconventional superconductivity with pair scattering between different Fermi surfaces. Since this necessarily involves spin-flip scattering, it suggests magnon-mediated pairing~\cite{Rohling2018_Superconductivity, Fjaerbu2019_Superconductivity, Kargarian2016_Amperean, Erlandsen2020_MagnonMediated, Hugdal2018_Magnon, Erlandsen2019_Enhancement, Thingstad2021_Eliashberg, Maeland2023_Skyrmionic, Maeland2023_Helical, Brekke2023_Twisted}. For the inversion symmetric case, the simplest pairing interaction does not conserve angular momentum, leaving preservation of $\mathcal{T} \otimes \sigma_h$ as the most viable option.

Group-VI TMD monolayers form hexagonal lattices with $D_{3h}$ point-group symmetry~\cite{liu2013}. 
They have broken in-plane-inversion symmetry, but preserve both \(\mathcal{T}\) and $\sigma_h$.  
This results in Ising spin-orbit coupling, such that the valence-band maxima occur with opposite out-of-plane spin at the two inequivalent valleys $K$ and $K'$. To break $\mathcal{T}$ symmetry in such a bilayer system while preserving $\mathcal{T}\otimes\sigma_h$, we need opposite Zeeman fields in the two layers. In addition, we need magnons to provide a spin-flip pairing mechanism, and an insulating barrier to prevent layer hybridization. All three can be provided by an intermediate hexagonal N\'eel antiferromagnetic insulator.

\emph{Heterostructure model.---}
We consider the heterostructure in Fig.~\ref{fig:heterostruct}, consisting of a honeycomb lattice antiferromagnetic insulator sandwiched between two TMD layers at low hole doping. The lattice of the AFMI can be partitioned into two triangular sublattices $A$ and $B$. The low-energy valence bands of TMDs have orbital characters dominated by the $d_{xy}$ and $d_{x^2-y^2}$ orbitals of the transition-metal atoms, which form a triangular lattice.  In our structure, the triangular lattice of the top (bottom) TMD is commensurate with and on top of (below) the sublattice composed of $A$ ($B$) sites of the AFMI.
Assuming a nearest neighbour exchange interaction between the itinerant electron spins in the TMDs and the local spins of the AFMI, we model this system by the Hamiltonian \(H=H_{\rm TMD}+H_{\rm AFMI}+H_{\rm int}\), where
\begin{eqnarray}
H_{\rm TMD}&=&\sum_{l\sigma}\sum_{i,j\in l} [ t^\sigma_{ij} - \delta_{ij} (\mu - \epsilon_0) ] c^\dagger_{il\sigma}c_{jl\sigma}, \label{eq:HTMD} \\
H_{\rm AFMI}&=& \sum_{m \alpha} \sum_{{\langle i, j \rangle}_m} J_{m}^\alpha S_i^\alpha S_j^\alpha-K\sum_i(S_i^z)^2, \\
H_{\rm int}&=&\bar{J}\sum_{l\sigma\sigma'}\sum_{i\in l}\b{S}_i\cdot\b{\sigma}^{\sigma\sigma'}c^\dagger_{il\sigma}c_{il\sigma'}. \label{eq_Hint}
\end{eqnarray}
Here,  $c^\dagger_{il\sigma}$ ($c_{il\sigma}$) creates (annihilates) an electron with spin $\sigma$ on the hexagonal lattice site $i$ of layer $l$, and $S_i^\alpha$ denotes the $\alpha$-component of the spin-$S$ operator at site $i$ in the AFMI, while $\b{\sigma}$ is a vector of Pauli matrices.  Here, ${\langle i, j \rangle}_m$ in the AFMI Hamiltonian denotes the sum over pairs of $m$th nearest neighbour sites $(i,j)$ on the honeycomb lattice.
From the stacking geometry, the index $i$ also correspond to sites of sublattice $l \in \{A, B\}$ of the AFMI, and we use the single index \(i\) to refer to both.
We assume that the hopping amplitudes \(t_{ij}^\sigma\) are non-zero only for nearest neighbour sites \(i,j\) on the triangular lattice (next-nearest neighbour on the hexagonal lattice). The symmetries of the TMDs constrain the hopping elements to the form \(t_{ij}^\sigma = t_0 + i \sigma \zeta_{ij} t \), 
with sign  \(\zeta_{ij} =  \pm 1\) determined by the hopping direction as shown in Fig.~\ref{fig:heterostruct}(c).
The hopping amplitudes $t_0$ and $t$ can be fit to ab-initio results for a given TMD material, as discussed in the Supplemental Material (SM)~\cite{suppMatShort}. For $\mathrm{MoSe_2}$, we obtain \(t_0 = \SI{-0.21}{\electronvolt}\) and \(t = \SI{-0.28}{\electronvolt}\), which we use throughout the paper.  Furthermore, \(\mu\) is the chemical potential, \(\epsilon_0 = -3 ( |t_0 | + \sqrt{3} |t|)\)~\footnote{The constant \(\epsilon_0\) shifts the spectrum so that the energy is zero on top of the valence band}, and $J_{m}^\alpha$ are the exchange coupling strengths, where we assume $J_m^{x} = J_m^y \equiv J_m$ throughout. Finally, $K$ is the easy axis anisotropy and \(\bar{J}\) the s-d exchange coupling.

For simplicity we consider spin-spin interactions up to nearest neighbours in the AFMI Heisenberg Hamiltonian. Motivated by $\mathrm{MnPSe_3}$~\cite{thuc2021, wildes1998, calder2021}, we use \(J_{1} =  J_1^z = \SI{0.9}{\milli\electronvolt}\) and \(S=5/2\). We further set the easy axis anisotropy to   \(K = 0.14  J_1\), although it could also be larger due to the strong spin-orbit coupling in the heterostructure. 
Applying a linearized Holstein-Primakoff transformation,
we write the magnon Hamiltonian in terms of the sublattice Fourier modes $a_\b{q}$ and $b_{\b{q}}$:
\begin{eqnarray}
H_{\rm AFMI} &=&\sum_{\b{q}} 
\begin{pmatrix}  a_{\b{q}}^\dagger & b_{-\b{q}}  \end{pmatrix}
\begin{pmatrix} C_\b{q} & D_\b{q} \\ D_\b{q}^* & C_\b{q} \end{pmatrix} 
\begin{pmatrix} a_\b{q} \\ b_{-\b{q}}^\dagger \end{pmatrix},
\label{eq:HAFMI}
\end{eqnarray}
where \(C_{\bm{q}}\equiv 3J_1^zS+2KS\) and \(D_{\bm{q}}\equiv J_1 S\gamma_{\b{q}}\). 
Here, $\gamma_{\b{q}}=\sum_{j} e^{i\b{q}\cdot\b{\delta}_j}$, and \(\bm{\delta}_j\)
are the nearest neighbour vectors from the $A$ to the $B$ sublattice of the AFMI. 
The magnon spectrum is obtained by directly diagonalizing  $H_{\rm AFMI}$, however, it also gets renormalized through the coupling to the electronic system. Incorporating this through the static polarization \(\Pi^\sigma_{\bm{q}}\) (see SM~\cite{suppMatShort}),
the effective frequency for a magnon of spin $\sigma$ is
\be
\omega^\sigma_{\b{q}}=\sqrt{( C_{\b{q}} + 2 S \bar{J}^2 \Pi^\sigma_{\bm{q}} )^2-|D_{\b{q}}|^2 },
\ee
where we set \(\hbar=1\). The bare (\(\Pi^\sigma_{\bm{q}}=0\)) and renormalized spectra are shown in Fig~\ref{fig:heterostruct}(e).

\begin{figure}[tbp]
\includegraphics[width=0.95\columnwidth]{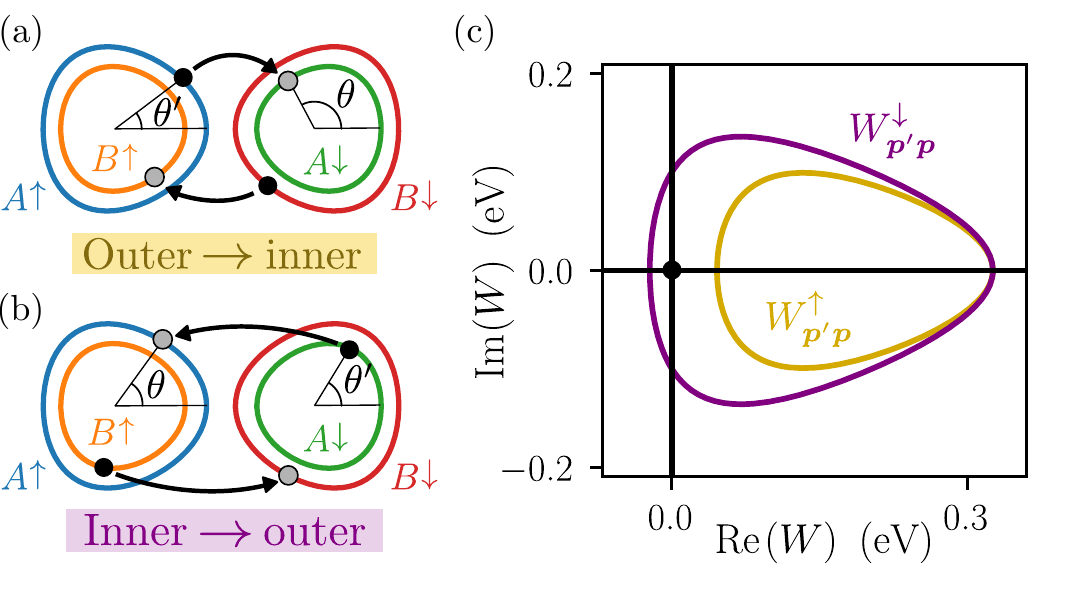}
\caption{(a) Schematic Fermi surfaces for the TMDs. The magnon-mediated pair scattering processes are indicated with arrows for the scattering of a Cooper pair from the outer to the inner Fermi surfaces.  
(b) Pair scattering process from the inner to the outer Fermi surfaces.  
(c) Pair scattering potential in the complex plane. The curves correspond to scattering from fixed incoming momentum \(\bm{p}'\) specified by \(\theta'\) to outgoing momentum \(\b{p}\) specified by \(\theta\) for the outer-to-inner (yellow; \(\theta'=0\)) and inner-to-outer (purple; \(\theta'=\pi\)).  
The latter encloses the origin, while the former does not. 
This is the reason for the coexisting topologically trivial and non-trivial gaps.
}
\label{fig:fig2}
\end{figure}

\emph{Effective interaction.---}
Superconductivity can arise through a magnon-mediated effective interaction due to the coupling term in Eq.~\eqref{eq_Hint}. Here, we consider the spin-wave expansion up to first order in the magnon operators, and write \(H_\mathrm{int} = H_\mathrm{int}^0 + H_\mathrm{int}^1\). The zeroth order term describes the Zeeman splitting experienced by the electrons due to the antiferromagnet and is included directly in the electronic Hamiltonian
\begin{align}
H_\mathrm{TMD} + H_\mathrm{int}^0 =  \sum_{\bm{k} l\sigma} \xi_{\b{k}l\sigma} c_{\bm{k} l \sigma}^\dagger c_{\bm{k} l\sigma},
\end{align}
where \(\xi_{\b{k}l\sigma} = \epsilon_{\b{k} l \sigma} - \mu\) and the spectrum is 
\begin{align}
\epsilon_{\b{k}l\sigma} = 
\epsilon_0 + 
2 \sum_n [t_0\cos(\b{k}\cdot\b{b}_n)+t\sigma\sin(\b{k}\cdot\b{b}_n)] + l \sigma \bar{J} S.
\label{eq_electronSpectrum}
\end{align}
Here, $\{\b{b}_n\}_{n=1}^3$ are the three triangular lattice nearest neighbour vectors in Fig.~\ref{fig:heterostruct}(c), and
$l=+/-$ for layer $A/B$. The spectrum is illustrated in Fig.~\ref{fig:heterostruct}(b) and gives rise to the schematic Fermi surfaces labelled by \(l,\sigma\) in Fig.~\ref{fig:fig2}(a).  

The first order term takes the form
\begin{align}
H_\mathrm{int}^1 =   \frac{\sqrt{2S} \bar{J}}{\sqrt{N}} \sum_l \sum_{\b{k},\b{q}} M_{\b{q}}^{l\sigma} c_{\b{k}+\b{q},l\sigma}^\dagger c_{\b{k}l\bar{\sigma}} ,
\end{align}
where we have introduced the magnon operators \(M_{\b{q}}^{l\sigma}\) given by \(M_{\b{q}}^{A\downarrow} = (M_{-\b{q}}^{A\uparrow})^\dagger = a_{\b{q}}\) and \(M_{\b{q}}^{B\uparrow} = (M_{-\b{q}}^{B\downarrow})^\dagger = b_{\b{q}}\), and \(N\) is the number of unit cells. 

The magnons can now be integrated out through a Schrieffer-Wolff transformation, to give rise to an effective interaction Hamiltonian.  
Due to the Fermi surface splitting,
on-shell superconducting pairing can only form between particles on the Fermi surfaces \((A\uparrow, B\downarrow)\) and on \((A\downarrow, B \uparrow)\). Furthermore, since the magnons carry a spin, the scattering processes mediating the pairing must be of inter-valley type, as also shown in Fig.~\ref{fig:fig2}(a-b). The relevant BCS reduced Hamiltonian for the expected superconducting pairing is therefore of the form
\begin{align}
H^\mathrm{BCS} &= \frac{1}{N} \sum_{\b{p}\b{p}' \sigma}
W_{\b{p}\b{p}'}^\sigma c_{\b{p}' A \bar{\sigma}}^\dagger c_{-\b{p}' B \sigma}^\dagger c_{-\b{p} B \bar{\sigma}} c_{\b{p} A \sigma} ,
\end{align}
\noindent with pair scattering potential
\begin{align}
W_{\b{p}\b{p}'}^\sigma = - 2 S \bar{J}^2   \frac{D_{\b{p}'-\b{p}}} {(\xi_{\b{p}A\sigma} - \xi_{\b{p}'A \bar{\sigma}})^2 - (\omega^{\bar{\sigma}}_{\b{p}'-\b{p}})^2 }.
\label{eq_pairScatteringPotential}
\end{align}
Notably, it is proportional to the sublattice hybridization \(D_{\bm{p}'-\bm{p}}\) in Eq.~\eqref{eq:HAFMI}. 
This is because the pair-scattering process occurs between electrons in different layers. Since the two layers are coupled to different sublattices of the antiferromagnet, the magnons mediating the interaction propagate from one sublattice to the other, so that the corresponding propagator is proportional to \(D_{\bm{q}}\), which is a complex number. Writing \(\bm{q} = \pm \bm{K} + \bm{\kappa}\) and expanding to linear order in the deviations \(\bm{\kappa}\), it takes the form \(D_{\pm \bm{K} + \bm{\kappa}} \propto (\kappa_x \pm i \kappa_y)\). Thus, its phase winds when the momentum \(\bm{q}\) is moved around the point \(\bm{K}\). The effect of this on the pair scattering potential can be observed by plotting \(W_{\bm{p}' \bm{p}}^\sigma\) in the complex plane, as shown in Fig.~\ref{fig:fig2}(c). The pair-scattering from the inner to outer Fermi surfaces results in a phase winding of the interaction potential, while scattering from the outer to inner Fermi surfaces does not. 
As we will see, this gives rise to two coexisting gap components: one topologically trivial, and one non-trivial. 

\emph{Superconductivity.---}
The gap equation can be derived by performing a standard mean-field decoupling of the BCS reduced Hamiltonian.
Introducing the superconducting gap function
\begin{align}
\Delta_{\bm{p}}^{A\sigma} \equiv  \frac{1}{N} \sum_{\bm{p}'} 
W_{\bm{p}'\bm{p}}^{\bar{\sigma}} \langle c_{-\bm{p}'{B} \sigma} c_{\bm{p}'{A} \bar{\sigma}} \rangle,
\end{align}
we obtain
\begin{align}
\begin{pmatrix} \Delta_{\b{p}}^{A\uparrow} \\ \Delta_{\b{p}}^{A\downarrow}   \end{pmatrix} 
= -  \frac{1}{N} \sum_{\b{p}'}  \begin{pmatrix} 0 
& W^{\downarrow}_{\b{p}'\b{p}} \chi_{\b{p}'}^{A\downarrow} 
\\ W^\uparrow_{\bm{p}'\bm{p}} \chi_{\b{p}'}^{A\uparrow} 
& 0 \end{pmatrix}  
\begin{pmatrix} \Delta_{\b{p}'}^{A \uparrow}  \\  \Delta_{\b{p}'}^{A\downarrow} \end{pmatrix},\label{eq:gapeqn}
\end{align}
\noindent with susceptibilities \(\chi_{\b{p}}^{l\sigma} = \tanh (\beta E_{\b{p}}^{l\sigma} /2)/2E_{\b{p}}^{l\sigma}\) and energies $E_{\b{p}}^{l\sigma} = \sqrt{\xi_{\b{p} l \sigma }^2 + | \Delta_{\b{p}}^{l\sigma} |^2 }$.
\noindent As expected from the relevant scattering processes, the gap equation is off-diagonal in the pairing amplitudes on the inner and outer Fermi surfaces. While the above gap equations determine \(\Delta_{\bm{p}}^{A\sigma}\), the gap function \(\Delta_{\bm{p}}^{B\sigma}\) on Fermi surface $B\sigma$ is obtained from the symmetry relation \(\Delta_{\bm{p}}^{ l \sigma} = - \Delta_{-\bm{p}}^{ \bar{l} \bar{\sigma}}\).
Due to the suppression of the pairing potential for momenta far away from the Fermi surface, we may restrict our attention to an energy range \(\omega_D \equiv \omega_{\b{K}}^\downarrow = \omega_{-\b{K}}^\uparrow\) around the Fermi level. By integrating out the perpendicular momentum, the gap equation can then be reduced to a gap equation on the Fermi surface (see SM~\cite{suppMatShort}).

\begin{figure}[tbp]
\includegraphics[width=0.95\columnwidth]{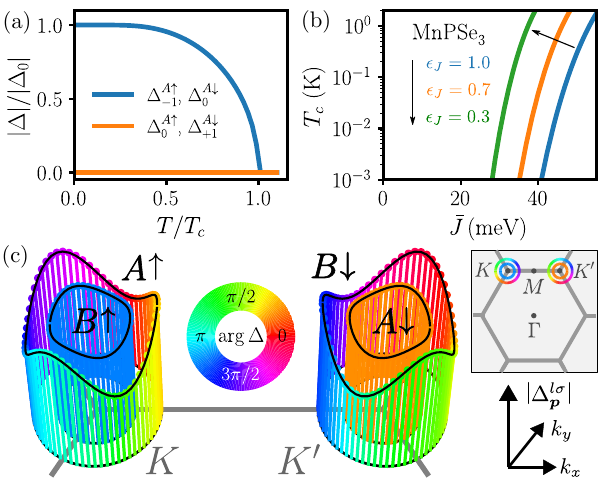}
\caption{
(a) Absolute value of the gap amplitudes determined from the gap equation in the limit of small hole doping. 
(b) Critical temperature $T_c$ as a function of interlayer exchange coupling obtained from the gap equation for varying biaxial strain (modelled by $J_1 =  \epsilon_J \times \SI{0.9}{\milli\electronvolt}$). 
(c) $\Delta_{\b{p}}^{l\sigma}$ on the four Fermi surfaces, as determined from the gap equation (Eq. \eqref{eq:gapeqn}) on the Fermi surface at \(T=0\). The height indicates the magnitude, and the color indicates the phase.}
\label{fig:fig3}
\end{figure}

Consider first the limit of small hole doping, resulting in small circular Fermi surfaces characterized by Fermi momenta  \(k_F^{A\sigma}\). 
At the critical temperature, the gap equation can be linearized, and utilizing the approximate form of \(D_{\pm \bm{K} + \bm{\kappa}}\), we obtain 

\begin{align}
\lambda \begin{pmatrix} \Delta^{A\uparrow}(\theta) \\ \Delta^{A\downarrow}(\theta) \end{pmatrix}
=
-\frac{v}{k_F} \int \frac{d\theta'} {2\pi} M(\theta, \theta') 
\begin{pmatrix}\Delta^{A\uparrow}(\theta') \\ \Delta^{A\downarrow}(\theta') \end{pmatrix},
\end{align}
where the Fermi surfaces have been parametrized by angle \(\theta\), see Fig. \ref{fig:fig2}. The effective pairing potential strength is
\begin{equation} \label{eq_effectiveCouplingStrengthAnalytical}
v =  \frac{3}{2} S^2  (k_F a)  \nu_F a^2  \frac{\bar{J}^2 J_1 }{\omega_D^2 } ,     
\end{equation}
where \(k_F  \equiv \frac{1}{2}(k_F^{A\uparrow}+ k_F^{A\downarrow})\) is the average Fermi wavenumber and \(\nu_F\) the density of states per spin at the Fermi level for a single TMD. 
Furthermore, \(M(\theta, \theta')\) is the matrix
\begin{align}
M(\theta, \theta') =  \begin{pmatrix}
0 & k_F^{A\uparrow} e^{-i \theta} - k_F^{A\downarrow} e^{-i \theta'} \\
k_F^{A\uparrow} e^{i \theta'} - k_F^{A\downarrow} e^{i \theta} & 0
\end{pmatrix}.
\end{align}
The above gap equation is an eigenvalue problem for the effective coupling strength \(\lambda\), which is related to the critical temperature \(T_c\) through \(\lambda^{-1} = \log \left(\frac{2 e^\gamma  \omega_D}{ \pi k_B T_c}\right) \), where \(\gamma\) is the Euler-Mascheroni constant. 
The angular dependence of \(M(\theta, \theta')\) fixes the angular dependence of the gap functions, which can be decomposed as \(\Delta^{A\uparrow}(\theta) = \Delta_0^{A\uparrow} + \Delta_{-1}^{A\uparrow} e^{-i \theta}\) and \(\Delta^{A\downarrow}(\theta) = \Delta_0^{A\downarrow} + \Delta_{+1}^{A\downarrow} e^{+i \theta}\). 
Inserting the ansatz, the gap equation splits into two separate pairs of equations, one for \((\Delta_{-1}^{A\uparrow}, \Delta_0^{A\downarrow } )\), and one for \((\Delta_0^{A\uparrow }, \Delta_{+1}^{A\downarrow})\). The leading instability occurs in the channel with the largest critical temperature, as obtained from the largest  eigenvalue \(\lambda = v k_F^{A\uparrow} /k_F\), where we assumed \(k_F^{A\uparrow} > k_{F}^{A\downarrow}\). The order parameter takes the form \(\bm{\Delta} = (\Delta_{-1}^{A\uparrow } e^{-i \theta}, \Delta_0^{A\downarrow} )^T \).  
Since the two pairs of gap components mutually suppress each other, the other pair \((\Delta_0^{A\uparrow}, \Delta_{+1}^{A\downarrow})\) remains zero all the way down to zero temperature. This is confirmed by the numerical solution of the gap equation at small hole doping, as shown in Fig.~\ref{fig:fig3}(a). 

We also solve the gap equation for larger doping, where the Fermi surfaces have trigonal warping. The resulting critical temperature is shown in Fig.~\ref{fig:fig3}(b) for exchange parameters motivated by $\mathrm{MnPSe_3}$.  
Biaxial strain can reduce the intralayer exchange interaction in the magnets significantly~\cite{sivadas2015magnetic, Xue2020_Control}, and we incorporate this by reducing all intralayer exchange parameters by a factor $\epsilon_J$. This enhances the critical temperature, as also shown in the figure. The gap profile on the Fermi surface at zero temperature is shown in Fig.~\ref{fig:fig3}(c). 
As expected from the solution in the limit of small hole doping, the order parameters \(\Delta_{\b{p}}^{A\uparrow}\) and \(\Delta_{\b{p}}^{B\downarrow}\) have chiral $p$-wave character, while \(\Delta_{\b{p}}^{ A \downarrow}\) and \(\Delta_{\b{p}}^{B\uparrow}\) have $s$-wave character. In addition, the gap develops amplitude modulations for non-circular Fermi surfaces~\footnote{The gap amplitude modulations are somewhat larger for the chiral $p$-wave component than for the $s$-wave component.}.

\begin{figure}[tbp]
\includegraphics[width=0.95\columnwidth]{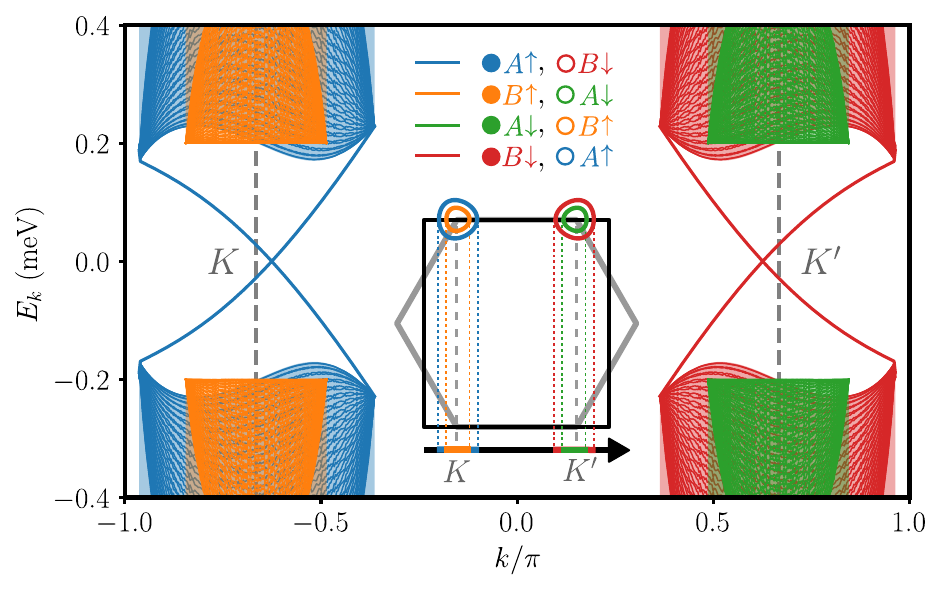}
\caption{Low-energy eigenstates in a ribbon geometry with \(N_y = 10^5\) sites and periodic boundary conditions in the $x$-direction. 
The eigenstates are composed of electrons from band \((l,\sigma)\) and holes from band \((\bar{l},\bar{\sigma})\). 
Two of the four gap components have chiral $p$-wave character, yielding two topological sectors. Each of these sectors hosts two Majorana edge modes (one on each edge).
The shaded regions indicate the projection of the bulk band. The inset shows the projection of the Brillouin zone down on the \(k_x\)-axis. 
We used chemical potential $\mu = - \SI{0.3}{\electronvolt}\) and  average gap magnitude $\SI{0.2}{\milli\electronvolt}$ on the Fermi surface.
}
\label{fig:fig4}
\end{figure}

\emph{Topological edge states.---}
To understand the topology of the system, we now consider the Bogoliubov-de Gennes (BdG) Hamiltonian with superconducting $s$-wave and chiral $p$-wave gap functions. It
takes the block diagonal form 
\begin{align}
H = \frac{1}{2} \sum_{\bm{p} l\sigma} 
\psi_{\bm{p} l \sigma}^\dagger 
\begin{pmatrix} \xi_{\bm{p} l \sigma} & \Delta_{\bm{p}}^{l\sigma} \\
\Delta^{l \sigma*}_{\bm{p}} & - \xi_{-\bm{p} \bar{l} \bar{ \sigma}} \end{pmatrix}
\psi_{\bm{p} l \sigma},
\label{eq_BdG}
\end{align}
where \(\psi_{\bm{p} l \sigma} = \begin{pmatrix} c_{\bm{p} l \sigma}, & c_{-\bm{p} \bar{l} \bar{\sigma}}^\dagger \end{pmatrix}^T\).
Two blocks
have a gap with extended $s$-wave character, while two blocks have a gap with chiral $p$-wave character. Calculating their Chern numbers~\cite{suppMatShort}, 
we find two topologically trivial blocks with Chern number \(0\), and two non-trivial blocks with Chern number \(+1\).
From the bulk boundary correspondence, we therefore expect topological edge modes in a finite geometry. 
To check this,
we compute the BdG spectrum on a ribbon geometry (see SM~\cite{suppMatShort}). We obtain the result in Fig.~\ref{fig:fig4}, showing edge states crossing the bulk bandgap for the topological \(p\)-wave blocks. These are chiral Majorana edge modes,
and they should also occur in vortices~\cite{Alicea2012_NewDirectionsPursuit}.  
We also note that a domain wall in the antiferromagnet would yield regions of pairing with opposite chirality, and therefore confine two Majorana edge modes
which can transport charge~\cite{serban2010}.

\emph{Discussion.---}
The critical temperature is determined by the effective coupling strength \(\lambda\), which depends on model parameters through~Eq.~\eqref{eq_effectiveCouplingStrengthAnalytical}. 
This guides the search for materials to realize the proposed phenomena~\cite{liu2023, Mak2019_Probing, Shi2023_RecentProgressInStrong}. 
In particular, 
the magnon frequency of the pair scattering processes should be small.
The interlayer exchange interaction \(\bar{J}\) should also be large. There is a growing list of heterostructures 
where this is the case~\cite{Zhai2019_SpinDependent, Yang2013_ProximityEffects, zhong2017, wu2020, qi2015, Scharf2017_Magnetic, Hallal2017_Tailoring, Gan2016_TwoDim, Qiao2014_Quantum, norden2019, Xu2018_Large, Wang2023_GiantSpontaneous, Lin2020_MagneticProximity, Li2018_LargeValleyPolarization}. 
For instance, graphene deposited on antiferromagnetic CrSe exhibits an induced splitting of \SI{134}{\milli\electronvolt}~\cite{norden2019}.  
Enormous proximity induced spin-splitting can also be achieved in TMDs~\cite{zhong2017, Xu2018_Large}, particularly when deposited on EuS and EuO ($\sim$\SI{300}{\milli\electronvolt})~\cite{wu2020, qi2015}. 
Furthermore, $\bar{J}$ can be boosted by hydrostatic pressure~\cite{zhang2018, fulop2021}. 
These examples show that reasonably large values of $T_c$ should be accessible. 

There are several potential material candidates for our structure, such as the experimentally studied $\mathrm{MoSe_2}$ coupled to the planar AFMI $\mathrm{MnPSe_3}$~\cite{Onga2020_AfmiSemicond} or the very similar $\mathrm{MnPS_3}$. These magnets exhibit a soft magnon spectrum, with excitation energy around \SI{10}{\milli\electronvolt} at the $K$-point~\cite{thuc2021, wildes1998}. The lattice spacing of the former is close to double that of $\mathrm{MoSe_2}$~\footnote{The electron spectrum therefore needs to be folded in on the Brillouin zone of the AFMI, but the only effect is an effective reduction of the exchange coupling strength.},
with a lattice mismatch of only \SI{2}{\percent}.
Since the emerging Moir\'e scale is large, one can expect puddles with topological interlayer pairing surrounded by normal regions. 
An alternative which does not depend on lattice matching is the layered A-type AFMI MnBi$_2$Te$_4$~\cite{Mak2019_Probing, hu2023}.  
It consists of alternating ferromagnetic layers whose natural stacking orientation ensures that a bilayer exactly maps onto our monolayer AFMI model. 
The magnon spectrum has a maximum of only $\SI{3}{\milli\electronvolt}$~\cite{lujan2022, Li2021_Quasi2d, Mai2021_MagnonPhonon, calder2021}, yielding critical temperatures similar to those obtained with $\mathrm{MnPSe_3}$~\cite{suppMatShort}. This demonstrates that several heterostructures are promising candidates to host topological interlayer superconductivity.

\emph{Acknowledgements.---}
This project has received funding from the European Union’s Horizon 2020 Research and Innovation Program under Grant Agreement No. 862046 and under Grant Agreement No. 757725 (the ERC Starting Grant). This work was supported by the Swiss National Science Foundation.

\nocite{Hasan2010_Colloquium}
\nocite{bernevig2013topological}


%

\newpage 
\setcounter{page}{0}

\onecolumngrid
\clearpage\thispagestyle{empty}

\begin{center}
        \Large       {Supplemental Material for \\ ``Topological Interlayer Superconductivity in a van der Waals Heterostructure''}
        
        \vspace{0.4cm}
        \normalsize
        Even Thingstad, Joel Hutchinson, Daniel Loss, and Jelena Klinovaja
        \\
        \textit{Department of Physics, University of Basel, Klingelbergstrasse 82, CH-4056 Basel, Switzerland} 
		\\ \vspace{0.2cm}
		(Dated: \today) 
       \vspace{0.5cm}
\end{center}

\renewcommand{\theequation}{S\arabic{equation}}
\renewcommand{\thefigure}{S\arabic{figure}}
\renewcommand{\thetable}{S\arabic{table}}
\renewcommand{\thesection}{S\arabic{section}}

\setcounter{section}{0}

This supplemental material contains a discussion of the electron spectrum (Sec.~\ref{sec_electrons}), the magnon spectrum and magnon renormalization (Sec.~\ref{sec_magnons}), the gap equation on the Fermi surface (Sec.~\ref{sec_gapEqFs}), its analytical solution in the limit of small hole doping (Sec.~\ref{sec_smallHoleDoping}), numerical results (Sec.~\ref{sec_numericalResults}), and topological edge states (Sec.~\ref{sec_topology}). 

\section{Electron spectrum}\label{sec_electrons}

Lightly hole-doped monolayer TMDs have two spin-valley locked bands close to the Fermi surface, and can therefore be described by a two-band tight-binding model. We consider only nearest neighbour hopping terms. The symmetries of the TMDs restrict the allowed hopping terms. TMDs have broken in-plane inversion symmetry but preserve mirror symmetry through the layer plane. 
The $C_3$ symmetry of lattice ensures isotropic hopping amplitudes. Finally, the TMDs preserve time-reversal symmetry. Together, this implies that the nearest neighbour hopping model is of the form given in the main text, in terms of two hopping parameters $t_0$ and $t$. 

These parameters can be set by comparing the low-energy spectrum obtained from the single-orbital model in the main paper to the more complicated three-orbital model in Ref.~\cite{liu2013}. At low hole-doping, we may expand around the points $\bm{K}=(\frac{4\pi}{3a},0)$ and $\bm{K}'=-\bm{K}$. 
With $\b{k}=\bm{K}+\kappa(\cos{\theta},\sin{\theta})$, the expansion of the spectrum (Eq.~\eqref{eq_electronSpectrum} in the main text) is
\be
\epsilon_{\b{k}l\sigma} - l \sigma \bar{J} S =   \frac{3}{4}(t_0+\sqrt{3}\sigma t)(\kappa a)^2 - \frac{1}{8}(\sqrt{3}t_0-\sigma t)\cos(3\theta)(\kappa a)^3+\mathcal{O}(\kappa^4).\label{eq:epsilon}
\ee
Similarly expanding the spectrum of the three-orbital model for MoSe$_2$~\cite{liu2013} around 
$\bm{K}$ gives
\begin{subequations}
\begin{eqnarray}
E_{\uparrow} / (\SI{1}{\electronvolt}) &\approx& {\rm const}-0.52(\kappa a)^2+0.0088\cos(3\theta)(\kappa a)^3.
\end{eqnarray}
\end{subequations}
Comparing the two expansions, we can extract the parameters $t_0=$ \SI{-0.21}{\electronvolt} and $t=$ \SI{-0.28}{\electronvolt} for MoSe$_2$. The corresponding density of states is $\nu_F = - [ 3\pi(t_0+\sqrt{3}t)a^2]^{-1}$.

\section{Magnon spectrum}\label{sec_magnons}

The AFMI Hamiltonian is given in Eq.~\eqref{eq:HAFMI} of the main text, where we assume \(J_{m}^x = J_{m}^y \equiv J_{m}^{xy} \equiv J_{m}\). Through a standard Holstein-Primakoff transformation and linear spin wave theory up to quadratic order in magnon operators, we then obtain the AFMI Hamiltonian

\begin{align}
H_\mathrm{AFMI} = \sum_{\bm{q}}  \left[
C_{\bm{q}} (a_{\bm{q}}^\dagger a_{\bm{q}}  + b_{\bm{q}}^\dagger b_{\bm{q}} ) 
+ (D_{\bm{q}}^* a_{\bm{q}} b_{-\bm{q}} + D_{\bm{q}} a_{\bm{q}}^\dagger b_{-\bm{q}}^\dagger )
\right],
\end{align}
\noindent where the coefficients \(C_{\bm{q}}\) and \(D_{\bm{q}}\) are in general (i.e. without assuming finite range interactions $J_m^\alpha$) given by
\begin{subequations}
\begin{align}
C_{\b{q}} &= \sum_m \sum_{j} \theta_m J_m S \cos (\bm{q} \cdot \bm{\delta}_j^m ) + \sum_m  (-1)^{\theta_m} z_m J_m^z S + 2 K S, \\ 
D_{\b{q}} &= \sum_m \sum_{j} (1 - \theta_m) J_m S e^{i \bm{q}\cdot \bm{\delta}_j^m }. \label{eq_suppMat_DGeneral}
\end{align}
\end{subequations}
\noindent Here $z_m$ is the number of $m$-th nearest neighbours on the honeycomb lattice, and $\b{\delta}_j^m$ are $m$-th nearest neighbour vectors labelled by $j \in \{1, \dots, z_m\}$ from a lattice site on sublattice $A$ to a lattice site on sublattice $A$ or $B$ (depending on $m$). For the nearest neighbours ($m=1$),  they are given by
$\{\b{\delta}_j^1\}_{j=1}^{3}=\frac{a}{\sqrt{3}}\{
(\frac{\sqrt{3}}{2}, \frac{1}{2}),\; 
(-\frac{\sqrt{3}}{2}, \frac{1}{2}), \;
(0, -1)\}$, where $a$ is the next-to-nearest neighbour distance on the honeycomb lattice, i.e. the lattice constant of the triangular Bravais lattice. 
The prefactor \(\theta_m\) is defined such that \(\theta_m = 1\) when the $m$th nearest neighbours are on the same sublattice, and \(\theta_m = 0\) otherwise. The above result is in fact generic for the AFMI magnon Hamiltonian on any bipartite lattice. The bare magnon spectrum $\omega_{\bm{q}}$ can now be obtained through a standard Bogoliubov transformation, and the result is 

\be \omega_{\bm{q}} = \sqrt{ C_{\bm{q}}^2 - |D_{\bm{q}}|^2}. \ee

Two antiferromagnetic materials described by a spin Hamiltonian of the form given in the paper are MnPSe$_3$~\cite{Onga2020_AfmiSemicond, calder2021} and a bilayer of the layered A-type AFM $\mathrm{MnBi_2Te_4}$~\cite{wildes1998, lujan2022}. The exchange and anistropy parameters utilized to calculate the critical temperature for these materials are given in Table~\ref{tab:magnonParam}, where we further assume $J_m^z = J_m$.
As discussed in the main paper, biaxial strain can reduce the in-plane exchange coupling constants significantly. We include this effect by reducing all in-plane exchange coupling strengths with a factor $\epsilon_J$, as also shown in the table. For the planar magnet $\mathrm{MnPSe_3}$~\cite{sivadas2015magnetic}, this amounts to reducing all exchange coupling strengths by $\epsilon_J$. For the layered magnet $\mathrm{MnBi_2Te_4}$~\cite{Xue2020_Control}, it means that only exchange couplings between spins on the same sublattice (i.e. within the same layer) are reduced.

\begin{table}[tbp]
\caption{Exchange parameters for candidate magnets $\mathrm{MnPSe_3}$~\cite{calder2021} and $\mathrm{MnBi_2Te_4}$~\cite{lujan2022} used to calculate the critical temperature in the main paper. Biaxial strain is modelled by a multiplicative factor $\epsilon_J$ reducing all intralayer exchange coupling strengths, i.e. $J_1$ for $\mathrm{MnPSe_3}$~\cite{sivadas2015magnetic} and $(J_2, J_5, J_6)$ for $\mathrm{MnBi_2Te_4}$~\cite{Xue2020_Control}. }\label{tab:magnonParam}
\vspace{0.2cm}
\begin{tabular}{l@{\hskip 0.4in}c@{\hskip 0.2in}c@{\hskip 0.2in}c@{\hskip 0.2in}c@{\hskip 0.12in}c@{\hskip 0.12in}c@{\hskip 0.2in}c@{\hskip 0.2in}c}
\hline 
Magnet & $S$ & $K$ (meV)  & $J_1$ (meV) & $J_2$ (meV) & $J_3$  (meV) & $J_4$ (meV) & $J_5$ (meV) & $J_6$ (meV) \\
\hline
$\mathrm{MnPSe_3}$ ($\epsilon_J=1.0$)& 5/2 & 0.045 & 0.9  &  &  &  &   \\
$\mathrm{MnPSe_3}$ ($\epsilon_J=0.7$) & 5/2 & 0.045 &  0.63  &  &  &  &  \\
$\mathrm{MnPSe_3}$ ($\epsilon_J=0.4$) & 5/2 & 0.045 & 0.36  &  &  &  &  \\
$\mathrm{MnBi_2Te_4}$ ($\epsilon_J=1.0$) & 5/2&0.04& 0.022 & -0.12 & 0 & 0 &  0.0332 & -0.0092\\
$\mathrm{MnBi_2Te_4}$ ($\epsilon_J=0.7$)  & 5/2 &0.04& 0.022 & -0.084 & 0 & 0 & 0.02324& -0.00644 \\
$\mathrm{MnBi_2Te_4}$ ($\epsilon_J=0.3$)  & 5/2 &0.04& 0.022  & -0.036  &  0 & 0 & 0.00996 & -0.00276\\
\hline
\end{tabular}
\end{table}

The back-action of the electrons on the magnons can be included through magnon renormalization, as follows. 
We introduce the magnon spinors \(\phi_{\bm{q} \downarrow} = \begin{pmatrix} a_{\bm{q}}, & b_{-\bm{q}}^\dagger \end{pmatrix}^T\) and \(\phi_{\bm{q} \uparrow} = \begin{pmatrix} a_{-\bm{q}}^\dagger, &  b_{\bm{q}} \end{pmatrix}^T\). The operators in the spinor \(\phi_{\b{q}\sigma}\) carry spin \(\sigma\), and since spin along the quantization axis is conserved in our model, the magnon propagator is block diagonal in these operators.  We therefore define the Matsubara magnon propagator 
\begin{align}
\mathcal{D}^\sigma(\b{q}, i \nu_m) = - \int_0^\beta d\tau \; e^{i \nu_m \tau} \langle \mathcal{T} \phi_{\b{q}\sigma} (\tau) \phi_{\b{q}\sigma}^\dagger(0) \rangle.
\end{align}
In the absence of coupling to the electronic system, this gives bare magnon propagator
\begin{align}
\mathcal{D}^\sigma_0(\bm{q}, i \nu_m) 
= \frac{1}{(i\nu_m)^2 - \omega_\b{q}^2} 
\begin{pmatrix}
i \sigma \nu_m + C_\b{q} & - D_\b{q} \\ - D_\b{q}^* & - i \sigma \nu_m + C_q 
\end{pmatrix}.
\label{eq_barePropagator}
\end{align}
In the main text, we give the effective pairing potential in Eq.~\eqref{eq_pairScatteringPotential}, as obtained by  deriving the effective interaction through a Schrieffer-Wolff transformation.  It can also be obtained directly from the magnon propagator through
\begin{align}
W_{\b{p} \b{p}'}^\sigma = V^2 \mathcal{D}^{\bar{\sigma}; A B}(\b{p}' - \b{p}, i\nu_m \rightarrow  \xi_{\b{p}' A \bar{\sigma}} - \xi_{\b{p} A \sigma} ),
\label{eq_pairScatteringSubstitution}
\end{align}
where $V = \sqrt{2S} \bar{J}$.
Inserting the off-diagonal element of the bare propagator from Eq.~\eqref{eq_barePropagator} for $\mathcal{D}$ in Eq.~\eqref{eq_pairScatteringSubstitution} clearly gives the result in Eq.~\eqref{eq_pairScatteringPotential} of the main text in the absence of coupling to the electronic system. However, the above relation is more general, and magnon renormalization can be included by dressing the bare propagator through RPA resummation of electron-hole bubbles in the TMDs to obtain $\mathcal{D}$. We therefore calculate the polarization  
\begin{align}
\Pi^{l \bar{\sigma}}(\b{q}, i \nu_m)=\frac{1}{N}\sum_{\b{k}}\frac{n_F(\xi_{\b{k} {l\sigma}}) - n_F(\xi_{\b{k}+\bm{q}, l\bar{\sigma}})}{i \nu_m  + \xi_{\b{k} {l\sigma}}-\xi_{\b{k} +\b{q},l \bar{\sigma}}},
\label{eq:polarizationExpression}
\end{align}
\noindent where \(n_F\) is the Fermi distribution function. Exploiting the symmetry relation \(\xi_{\b{k}l\sigma} = \xi_{-\b{k} \bar{l} \bar{\sigma}}\), one may show that \(\Pi^{l\sigma}(\b{q}, i\nu_m) = \Pi^{\bar{l} \sigma}(\b{q}, - i\nu_m)\). For the static polarization \(\Pi^{l\sigma}_\b{q} \equiv \Pi^{l\sigma}(\b{q}, 0) \), this means \( \Pi_\b{q}^{\sigma} \equiv \Pi_{\b{q}}^{l\sigma} = \Pi_\b{q}^{\bar{l} \sigma} \).
The Dyson equation $(\mathcal{D}^\sigma)^{-1} = (\mathcal{D}_0^\sigma)^{-1} - V^2\Pi^\sigma$ gives
\begin{align}
\mathcal{D}^\sigma(\b{q}, i \nu_m) = \frac{1}{(i\nu_m)^2 - (\omega_\b{q}^\sigma)^2} \begin{pmatrix}  i \sigma \nu_m + C_\b{q} + V^2\Pi^\sigma_\b{q} & - D_\b{q} \\ - D^*_\b{q} & -i \sigma \nu_m + C_\b{q} + V^2\Pi^\sigma_\b{q} \end{pmatrix},
\end{align}
\noindent with renormalized spectrum \(\omega_\b{q}^\sigma = [( C_\b{q} + V^2\Pi_\b{q}^\sigma)^2 - | D_\b{q}|^2 ]^{1/2}\).

For scattering momenta close to $\bm{q} = 0$ and $\bm{q} = \pm \bm{K}$, the polarization can be related to the standard Lindhard function by assuming that the electron spectrum is parabolic and that $\delta_z\equiv\bar{J}S$ is small compared to the electronic energy scale.  
For processes with small momentum $\bm{q}$, the polarization is small, while for intervalley scattering momentum  $\b{q}=\sigma\b{K}+\delta\b{q}$, the polarization is
\begin{eqnarray}
\Pi^{\bar{\sigma}}(\b{q} = \sigma \b{K}+\delta\b{q}, 0)&\approx &-\nu_F A_\mathrm{uc} \left[1-\Theta(|\delta\b{q}|-2k_F)\sqrt{1-\left(\frac{2k_F}{|\delta\b{q}|}\right)^2}\right]
= -\nu_F A_\mathrm{uc}, 
\end{eqnarray}
where \(A_\mathrm{uc} = \sqrt{3} a^2 / 2\) is the real space unit cell area.

\section{Gap equation on the Fermi surface}\label{sec_gapEqFs}

The gap equation is given in Eq.~\eqref{eq:gapeqn} of the main paper.  
From the interaction potential, we expect that only the regions within an energy range corresponding to the magnon energy can contribute to the pairing. Thus, we may restrict the momentum \(\b{p}'\) to this narrow region around the Fermi surface. In analogy with BCS theory, in this region, we may approximate the pairing potential by 
\begin{align}
W_{\bm{p} \bm{p}'}^\sigma = 
\Theta( \omega_{\b{p}'-\b{p}}^{{\bar{\sigma}}} - |\xi_{\b{p} A \uparrow}| ) \Theta( \omega_{\b{p}'-\b{p}}^{{\bar{\sigma}}} - |\xi_{\b{p}' A \downarrow}|) 
\frac{ V^2 D_{\b{p}'-\b{p}}}{(\omega_{\b{p}'-\b{p}}^{{\bar{\sigma}}})^2} .
\end{align}
\noindent Since the region that contributes to the pairing $W_{\bm{p} \bm{p}'}^\sigma$ is narrow, the integral over momentum in the gap equation can be rewritten in terms of integrals over momentum components parallel ($p_\parallel$) and perpendicular ($p_\perp$) to the Fermi surface. Furthermore, $W_{\bm{p} \bm{p}'}^\sigma$  is only weakly dependent on the momentum $p_\perp$ perpendicular to the Fermi surface. As a consequence, 
$\Delta_\b{p}^{l \sigma}$ is also approximately independent of this momentum component. The susceptibility \(\chi^{A\bar{\sigma}}_{\bm{p}'}\) only depends on \(\b{p}'\) through the energy \(\xi_{\bm{p}' A \bar{\sigma}}\), and is sharply peaked near the Fermi surface. Thus, we may further rewrite the perpendicular momentum integral in terms of an energy integral up to the approximate magnon cutoff frequency \(\omega^\sigma_{p_\parallel-p'_\parallel}\). The perpendicular momentum component can then be integrated out. Introducing the Fermi velocity \(v_F^{A\bar{\sigma}}(p_\parallel) = |\nabla_{\b{p}} \xi_{\b{p} A \bar{\sigma}} |\) on  Fermi surface $(A\bar{\sigma})$, we obtain gap equation

\begin{align}
\Delta^{A\sigma }(p_\parallel) = - \frac{1}{A_\mathrm{BZ} } \int_{{\rm FS}_{A\bar{\sigma}}} \frac{d p'_\parallel}{v_F^{A\bar{\sigma}}(p'_\parallel)}     W_{p'_\parallel p_\parallel}^{\bar{\sigma}} g(\omega^\sigma_{p-p'},| \Delta^{A \bar{\sigma}}(p'_\parallel)|)  \Delta^{A\bar{\sigma}}(p'_\parallel),
\label{eq:gapFS}
\end{align}

\noindent where the function $g(\omega, \Delta)$ is given by
\begin{align}
g(\omega,\Delta) \equiv \int_0^\omega d\xi \, \frac{\tanh ( \beta \sqrt{\xi^2 + |\Delta|^2}/2)}{ \sqrt{\xi^2 + |\Delta|^2}},
\end{align}
and $\beta$ is the inverse temperature.
In the special cases $T=T_c$ and $T=0$, the function $g$ is given by 

\begin{align}
g(\omega, 0) = \log \left( \frac{2 e^{\gamma}}{\pi} \beta \omega \right), 
\qquad \qquad 
\lim_{\beta \rightarrow \infty} g(\omega, \Delta) = \operatorname{arcsinh} \left( \frac{\omega}{|\Delta| } \right) \approx \log \left(\frac{2\omega}{|\Delta|}\right),
\label{eq:suppMat_gLimitTc}
\end{align}
which simplifies the calculation of $T_c$ and the gap at zero temperature.
We have now reduced the two-dimensional gap equation to a gap equation on the Fermi surface contour.
In the paper, we solve the gap equation in Eq.~\eqref{eq:gapFS} numerically at $T=0$ and $T=T_c$, and analytically in the limit of small hole doping. 

\section{Analytical solution in the limit of small hole doping}\label{sec_smallHoleDoping}
In the main text, we derive an analytical solution of the gap equation by assuming nearest-neighbour exchange interactions in the magnet and small hole doping. 
The latter allows three simplifications. First, the electron energy bands are approximately parabolic and the Fermi surfaces are circular, parametrized by an angle \(\theta\). Second, since the magnon frequency remains nearly constant as we vary the momenta \(\b{p}\) and \(\b{p}'\) on the Fermi surfaces, we let \(\omega_{\b{p}-\b{p}'}^\sigma \rightarrow \omega_D \equiv \omega_{\b{K}}^\downarrow = \omega_{-\b{K}}^\uparrow\). Third,  we can expand the interaction potential in small deviations \(\b{\kappa}\) from the valley maxima at \(\pm\bm{K}= \pm (\frac{4 \pi}{3a} ) (1, 0)\). For this, we write
\begin{eqnarray}
\bm{p}= \sigma \bm{K} + \b{\kappa}, \qquad \bm{p}' = - \sigma \bm{K} +  \b{\kappa}', \label{eq:suppMat_p}
\end{eqnarray} 
where $\b{\kappa}\equiv k_F^{A\sigma} (\cos \theta, \sin \theta)$ and $\b{\kappa}'\equiv k_F^{A\bar{\sigma}} (\cos \theta', \sin \theta')$. Here, \(\b{p}\) is on the Fermi surface \(A\sigma\), and \(\bm{p}'\) is on the Fermi surface \(A \bar{\sigma}\). We can then write \( \Delta^{l\sigma}_{\bm{p}}=\Delta^{l\sigma}(\theta)\).
Assuming a constant magnon frequency, the scattering potential \(W^{\bar{\sigma}}_{\b{p}'\b{p}}\) depends on the magnon momentum \(\bm{p}' - \bm{p}\) only through the factor \(D_{\bm{p} - \bm{p}'}\). For electron scattering processes between points on the Fermi surfaces,
the momenta are given by Eq.~\eqref{eq:suppMat_p}, so that
\begin{eqnarray}
D_{\b{p}-\b{p}'} =  J_1^{xy} S \sum_j e^{i (2\sigma\bm{K} + \bm{\kappa}- \bm{\kappa}') \cdot \bm{\delta}_j^1} 
\simeq \frac{\sqrt{3}a}{2} J_1^{xy} S\sigma\left(k_F^{A\sigma}e^{-i\sigma\theta}-k_F^{A\bar{\sigma}}e^{-i\sigma\theta'}\right).
\end{eqnarray}
With this approximation, the gap equation takes the form
\begin{align}
\begin{pmatrix} \Delta^{A\uparrow}(\theta) \\ \Delta^{A\downarrow}(\theta) \end{pmatrix}
=
-\frac{v}{k_F} \int \frac{d\theta'} {2\pi} 
\begin{pmatrix}
0 & +k_F^{A\uparrow} e^{-i \theta} - k_F^{A\downarrow} e^{-i \theta'} \\
k_F^{A\uparrow} e^{i \theta'} -k_F^{A\downarrow} e^{i \theta}  & 0
\end{pmatrix}
\begin{pmatrix} g(\omega_D, | \Delta^{A \uparrow}(\theta')|)  \Delta^{A\uparrow}(\theta') \\ 
g( \omega_D, | \Delta^{A \downarrow}(\theta')|) \Delta^{A\downarrow}(\theta') \end{pmatrix},
\label{eq:gapEqSmallBeforeLinearization}
\end{align}
where \(k_F\) is the average Fermi wavevector \(k_F = (k_F^{A\uparrow} + k_F^{A\downarrow})/2\), and $v$ is the effective coupling strength defined in Eq.~\eqref{eq_effectiveCouplingStrengthAnalytical} of the main text. There, we also discuss how this equation is solved analytically for $T=T_c$.

\begin{figure}
\includegraphics[width=0.6\textwidth]{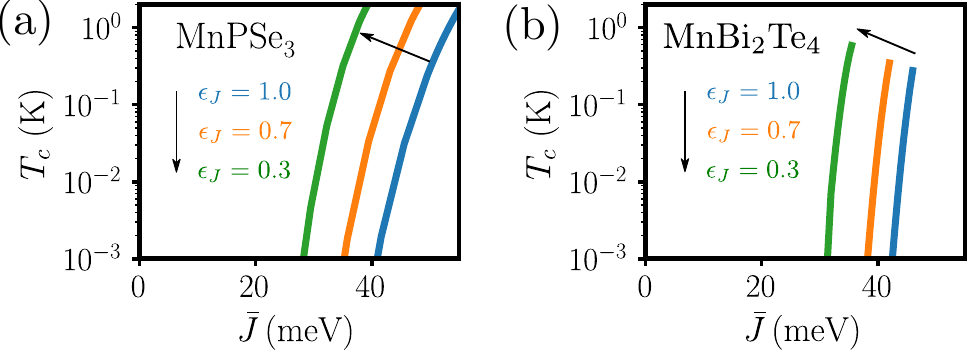}
\caption{Critical temperature obtained by solving the linearized gap equation in Eq.~\eqref{eq_gapEquationNumericalSolution} at finite hole doping. Exchange parameters as given in Tab.~\ref{tab:magnonParam} for (a)  MnPSe$_2$ (as also given in the main text) and (b) MnBi$_2$Te$_4$ for strained ($\epsilon_J = 0.7, 0.3$) and unstrained ($\epsilon_J = 1$) antiferromagnets. }
\label{fig:suppMat_gapProfileTc}
\end{figure}

\section{Numerical results at finite doping}\label{sec_numericalResults}

For \(T=T_c\), the gap equation can be linearized, and utilizing the result in Eq.~\eqref{eq:suppMat_gLimitTc}, we obtain 
\begin{align}
\Delta^{A\sigma}(\theta) = -2\pi \frac{V^2}{ A_\mathrm{BZ} }  \int_{\Gamma_{A\bar{\sigma}}} \frac{d \theta'}{2\pi} \frac{k_F^{A\bar{\sigma}}(\theta')}{v_F^{A\bar{\sigma}}(\theta') }  
\left( \frac{ D_{\b{p}-\b{p}'}}{(\omega^{{\sigma}}_{\b{p}-\b{p}'})^2}   \right)
\left[ 
\log \left( \frac{ \omega^{{\sigma}}_{\b{p} - \b{p}'}}{\bar{\omega}} \right) 
+ \log \left( \frac{2 e^\gamma}{\pi} \beta \bar{\omega} \right) 
\right]
\Delta^{A\bar{\sigma} }(\theta'),
\label{eq_gapEquationNumericalSolution}
\end{align}
\noindent where \(\b{p} = \sigma \bm{K} + k_F^{A\sigma} (\theta) (\cos \theta, \sin \theta)\), \(\b{p}' = - \sigma \b{K} + k_F^{A\bar{\sigma}}(\theta') (\cos \theta', \sin \theta')\), and \(\bar{\omega}\) is an arbitrary frequency scale. Computationally, however, it is advantageous to choose it to be a frequency that  is representative for the magnons mediating the pairing interaction. For simplicity, we therefore choose the magnon frequency averaged over the scattering processes involved in the pairing,
\begin{align}
\bar{\omega} = \langle \omega_{\b{q}}^\sigma \rangle_{\mathrm{proc}} \equiv \frac{1}{2} \sum_\sigma \int \frac{d\theta d\theta'}{(2\pi)^2} \langle \omega^{{\sigma}}_{\b{p} - \b{p}'} \rangle .
\end{align}
The above gap equation is now a generalized eigenvalue problem for the gap function with eigenvalue \(\lambda_c^{-1} \equiv \log \left( \frac{2 e^\gamma}{\pi} \beta \bar{\omega} \right) \). Solving it numerically, we thus obtain both the critical temperature \(T_c = \frac{2 e^\gamma}{\pi} \bar{\omega} \exp( - 1/\lambda_c)\) and the gap profile \(\Delta^{A\sigma}(\theta)\) for the leading instability.

In the main paper, we show the critical temperature as function of s-d exchange coupling $\bar{J}$ with exchange coupling constants corresponding to our model for MnPSe$_2$. We also calculate the critical temperature with exchange parameters corresponding to MnBi$_2$Te$_4$ as given in Table~\ref{tab:magnonParam} for different biaxial strain, and the result is shown in Fig.~\ref{fig:suppMat_gapProfileTc}.

\section{Topological edge states}\label{sec_topology}

\subsection{Topological invariant}

In terms of the Pauli matrix vector \(\b{\sigma}\), Eq.~\eqref{eq_BdG} of the main text can be written in the form 
\begin{align}
H = \frac{1}{2} \sum_{\b{k} l \sigma} \psi_{\b{k}l\sigma}^\dagger (\bm{d}_{\bm{k}}^{l\sigma} \cdot \bm{\sigma} ) \psi_{\b{k}l\sigma} ,\;\; \b{d}_{\bm{k}}^{l\sigma}=(\operatorname{Re}\Delta_{\b{k}}^{l\sigma}, -\operatorname{Im} \Delta_{\b{k}}^{l\sigma}, \xi_{\b{k}l\sigma}) .
\end{align}
For a Hamiltonian of this form, the Chern number is given by~\cite{Hasan2010_Colloquium}
\begin{align}
\mathcal{C}^{l\sigma} = \frac{1}{4\pi} \int_\mathrm{BZ} d^2{k} \,\, \bm{d}_{\bm{k}}^{l\sigma} \cdot (\partial_{k_x} \b{d}_{\bm{k}}^{l\sigma} \times \partial_{k_y} \b{d}_{\bm{k}}^{l\sigma} ).
\end{align}
The Chern number can subsequently be calculated numerically for the various blocks. For the blocks with chiral $p$-wave pairing, we find \(\mathcal{C}^{A\uparrow} = \mathcal{C}^{B\downarrow} = 1\). The chirality is identical because the phase winding of the gap runs in the same direction, as shown in Fig.~3(c) of the main text. For the blocks with $s$-wave pairing, we find \(\mathcal{C}^{A\downarrow} = \mathcal{C}^{B\uparrow} = 0\).

\subsection{Numerical calculation of the edge states}

To investigate possible topological edge states, we need a gap function defined throughout the Brillouin zone. Instead of using the solutions of the gap equation on the Fermi surface (Eq.~\eqref{eq:gapFS}) directly, we therefore use the topologically equivalent profiles 
\begin{align}
\Delta_{\b{p}}^{A\uparrow} = \Delta_\chi g_\chi(\bm{p} ), \qquad \Delta_{\b{p}}^{A\downarrow} = \Delta_s.
\end{align}
Here, \(g_\chi(\bm{p})\) is a basis function from the \(E_2\) irreducible representation of the symmetry group of the triangular lattice which reproduces the chiral $p$-wave gap on the Fermi surface. In particular, we choose 
\begin{align}
g_\chi(\bm{p} ) = \frac{2}{\sqrt{3}} \sum_n e^{i\phi_n} \cos (\bm{k} \cdot \bm{b}_n),
\end{align}
\noindent where \(e^{i\phi_n} = (\hat{x} + i \hat{y}) \cdot \b{b}_n / a\). 
The BdG Hamiltonian can now be rewritten in terms of a real-space model \(H=H_0 + H_1^s + H_1^\chi\) on the triangular lattice, where
\begin{subequations}
\begin{align}
H_0 &= \sum_{l\sigma}\sum_{i,j\in l} [ t^\sigma_{ij} - \delta_{ij} (\mu-l\sigma\bar{J}S - \epsilon_0)] c^\dagger_{il\sigma}c_{jl\sigma},
\\
H_1^s &= \Delta_s \sum_j (c_{j A \downarrow}^\dagger c_{jB \uparrow}^\dagger + c_{j B \uparrow} c_{j A \downarrow} ), \\
H_1^\chi &=  \frac{\Delta_\chi}{\sqrt{3}} \sum_{j n \zeta} \left( e^{i \phi_n } c_{j + \zeta \b{b}_n, A \uparrow}^\dagger c_{j B \downarrow}^\dagger + e^{-i\phi_n } c_{j B \downarrow} c_{j + \zeta \b{b}_n, A \uparrow} \right),
\end{align}
\end{subequations} with \(N_x \times N_y\) sites and periodic boundary conditions in the \(x\)-direction but not in the \(y\)-direction. To diagonalize the Hamiltonian, we introduce a standard partial Fourier transform and obtain the spectrum as a function of the momentum component $k_x$ along the $\hat{x}$-direction. The result is shown in Fig.~\ref{fig:fig4} of the main paper. 

\end{document}